\documentclass{article}

\usepackage{Sweave}
\usepackage{amsmath}
\usepackage{hyperref}
\usepackage{cite}
\usepackage{authblk}
\usepackage{graphicx}
\let\proglang=\textsf

\newtheorem{theorem}{Theorem}[section]
\newtheorem{lemma}[theorem]{Lemma}

\newcounter{algorithmctr}
\newenvironment{algorithm}{
   \refstepcounter{algorithmctr}
   \bigskip\noindent
   \textbf{Algorithm \thealgorithmctr\\}
}
{\bigskip}
\numberwithin{algorithmctr}{section}

\title{Efficient Thresholded Correlation using Truncated Singular Value Decomposition}
\author[1]{James Baglama \thanks{jbaglama@math.uri.edu}}
\author[2]{Michael Kane \thanks{michael.kane@yale.edu}}
\author[3]{Bryan Lewis \thanks{blewis@illposed.net}}
\author[3]{Alex Poliakov \thanks{apoliakov@paradigm4.com}}
\affil[1]{
University of Rhode Island
}
\affil[2]{
Yale University
}
\affil[3]{
Paradigm4, Inc.
}

\begin{document}
\maketitle
\begin{abstract}
Efficiently computing a subset of a correlation matrix consisting of values
above a specified threshold is important to many practical applications.
Real-world problems in genomics, machine learning, finance other applications
can produce correlation matrices too large to explicitly form and tractably
compute. Often, only values corresponding to highly-correlated vectors are of
interest, and those values typically make up a small fraction of the overall
correlation matrix. We present a method based on the singular value
decomposition (SVD) and its relationship to the data covariance structure that
can efficiently compute thresholded subsets of very large correlation matrices.
\end{abstract}

\section{Introduction}\label{intro}

Finding highly-correlated pairs among a large set of vectors is an important
part of many applications. For instance, subsets of highly-correlated vectors
of gene expression values may be used in the discovery of networks of genes
relevant to particular biological processes \cite{genomics}.  Identification of
highly-correlated pairs may be used in feature selection algorithms for machine
learning applications \cite{ml1, ml2} and are important to time series and
image processing applications \cite{timeseries, svd-similarity}.

The number of correlation coefficients grows quadratically with the number of
vectors. Simply computing all pairs of correlation coefficients and then
filtering out coefficients below a given threshold may not be computationally
feasible for high-dimensional data in modern genomics and other applications.
Considerable attention has been devoted to \emph{pruning methods} that cheaply
prune away all but the most highly-correlated pairs of vectors.  For example
the TAPER and related algorithms of Xiong et al. develop pruning rules based on
the frequency of mutually co-occurring observations (rows) among vector pairs;
see \cite{prune1, prune2} and the references therein.  Related methods
approximate a set containing the most highly-correlated pairs using hashing
algorithms \cite{prune3}. A number of other methods arrive at approximate sets
of the most highly-correlated pairs using dimensionality-reduction techniques
like the discrete Fourier transform \cite{timeseries}. The method of Wu, et al.
\cite{svd-similarity}, conceptually very similar to our approach, uses the
singular value decomposition (SVD) to find all pairs of close vectors with
respect to a distance metric (instead of correlation). Hero and
Rajaratnam\cite{hero2011large} apply projections to the thresholded correlation
problem with the additional objective of determining a reasonable threshold
value; their method follows a probabilistic approach.  Our method finds all of
the most highly-correlated vector pairs with respect to a given threshold by
pruning along an intuitive path of decreasing variance defined by the data and
computed by a truncated SVD.

Consider a real-valued data matrix $A$ consisting of $m$ observations of $n$
column vectors.  Denote the columns of $A$ as $a_j\in\mathcal{R}^m$ for
$j=1,2,\ldots,n$, $A=[a_1, a_2, \ldots, a_n]\in\mathcal{R}^{m\times n}$, and
let $k=\mathrm{rank}(A)$.  Assume that the mean of each column $a_j$ is zero
and that each column has unit norm $\|a_j\| = 1$. Here and below, $\|\cdot\|$
denotes the Euclidean vector norm. Then the Pearson sample correlation matrix
$\mathrm{cor}(A)=A^TA$.  Note that under these assumptions the sample
correlation and covariance matrices are the same up to a constant multiple.
Section \ref{irlba} illustrates how to relax the unit norm and zero mean
assumptions in practice.

Let $A=U\Sigma V^T$ be a singular value decomposition of $A$, where
$U\in\mathcal{R}^{m\times k}$, $V\in\mathcal{R}^{n\times k}$,
$U^TU = V^TV = I$,
and $\Sigma\in\mathcal{R}^{k\times k}$ is a diagonal matrix with
diagonal entries $\sigma_1 \ge \sigma_2 \ge \cdots \ge \sigma_k > 0$.  For
later convenience, we write $s_{1:p}=[\sigma_1, \sigma_2, \ldots, \sigma_p]^T$
to be the vector of the first $p\le k$ singular values along the diagonal of
$\Sigma$.  Denote the columns of $V$ as $V=[v_1, v_2, \ldots, v_k]$, where each
$v_j\in\mathcal{R}^n$, $j=1,2,\ldots,k$.  Then $\mathrm{cor}(A) = A^TA =
V\Sigma^2 V^T$ is a symmetric eigenvalue decomposition of the correlation
matrix with nonzero eigenvalues $\sigma_1^2, \sigma_2^2, \ldots, \sigma_k^2$.
The columns of $V$ form an orthonormal basis of the range of the
correlation matrix $A^TA$.

Let $a_i$ and $a_j$, $1\le i,j\le n$ be any two columns of the matrix $A$
(including possibly the case $i=j$). Denote the correlation of these two
vectors by $\mathrm{cor}(a_i,a_j) = a_j^T a_i$.  The following simple lemma
establishes a relationship between correlation and Euclidean distance between
two zero mean, unit norm vectors.

\begin{lemma}
\label{distlemma}
Let $a_i$ and $a_j$, $1\le i,j\le n$ be any two columns of the matrix
$A$. Then
\begin{equation}\label{cordist}
\mathrm{cor}(a_i,a_j) = a_i^Ta_j = 1 - \|a_i - a_j\|^2/2.
\end{equation}
In particular, for a given correlation threshold $t$, $0<t<1$, if
$\|a_i - a_j\|^2 > 2(1-t)$ then
$\mathrm{cor}(a_i, a_j) < t$.
\end{lemma}

Lemma \ref{distlemma} equates the problem of finding pairs of highly correlated
column vectors from the matrix $A$ with a problem of finding pairs of vectors
sufficiently close together. For example, correlation values larger than 0.99
may only be associated with column vectors from the matrix $A$ whose Euclidean
distance is less than $\sqrt{0.02}$.

The following lemma expresses the Euclidean distance between any two columns of
the matrix $A$ as a weighted sum of projected distances along the SVD
basis vectors forming the columns of $V$.
\begin{lemma}
\label{sumlemma}
Let $a_i$ and $a_j$
be any two columns of the matrix
$A$, $1\le i,j\le n$. Then
\begin{equation}\label{vdist}
\|a_i - a_j\|^2 =
\sigma_1^2 (e_i^Tv_{1} - e_j^Tv_{1})^2 + 
\sigma_2^2 (e_i^Tv_{2} - e_j^Tv_{2})^2 + \cdots + 
\sigma_k^2 (e_i^Tv_{k} - e_j^Tv_{k})^2,
\end{equation}
\end{lemma}
where, here and below, $e_{j}\in\mathcal{R}^n$ represents the $j$th unit basis
vector consisting of $1$ in position $j$ and zeros otherwise.
(Thus, $e_j^Tv_{k}$ is the $j$th position in the vector $v_k$--that is, the
$[j,k]$ entry in the matrix $V$.)


Note that each term in the sum is nonnegative and the weights $\sigma_j^2$,
$j=1,2,\ldots,k$ are nonincreasing and defined by the covariance structure of
the data.  Given a correlation threshold $t$, $0<t<1$, Equations \ref{cordist}
and \ref{vdist} together suggest that pairs of vectors can be excluded from
consideration whenever their projected distance is too large.  For example if
$\sigma_1^2 (e_i^Tv_{1} - e_j^Tv_{1})^2 > 2(1-t)$, then we can conclude from
just a few scalar values that the column vectors $a_i$ and $a_j$ do not meet the
correlation threshold $t$.
In practice, many pairs of vectors may be pruned in this fashion by
inspecting distances in low-dimensional subspaces, substantially reducing the
complexity of computing thresholded correlation matrices.

\section{Efficient pruning using truncated SVD}

\begin{figure}[!ht]
\begin{center}
\includegraphics{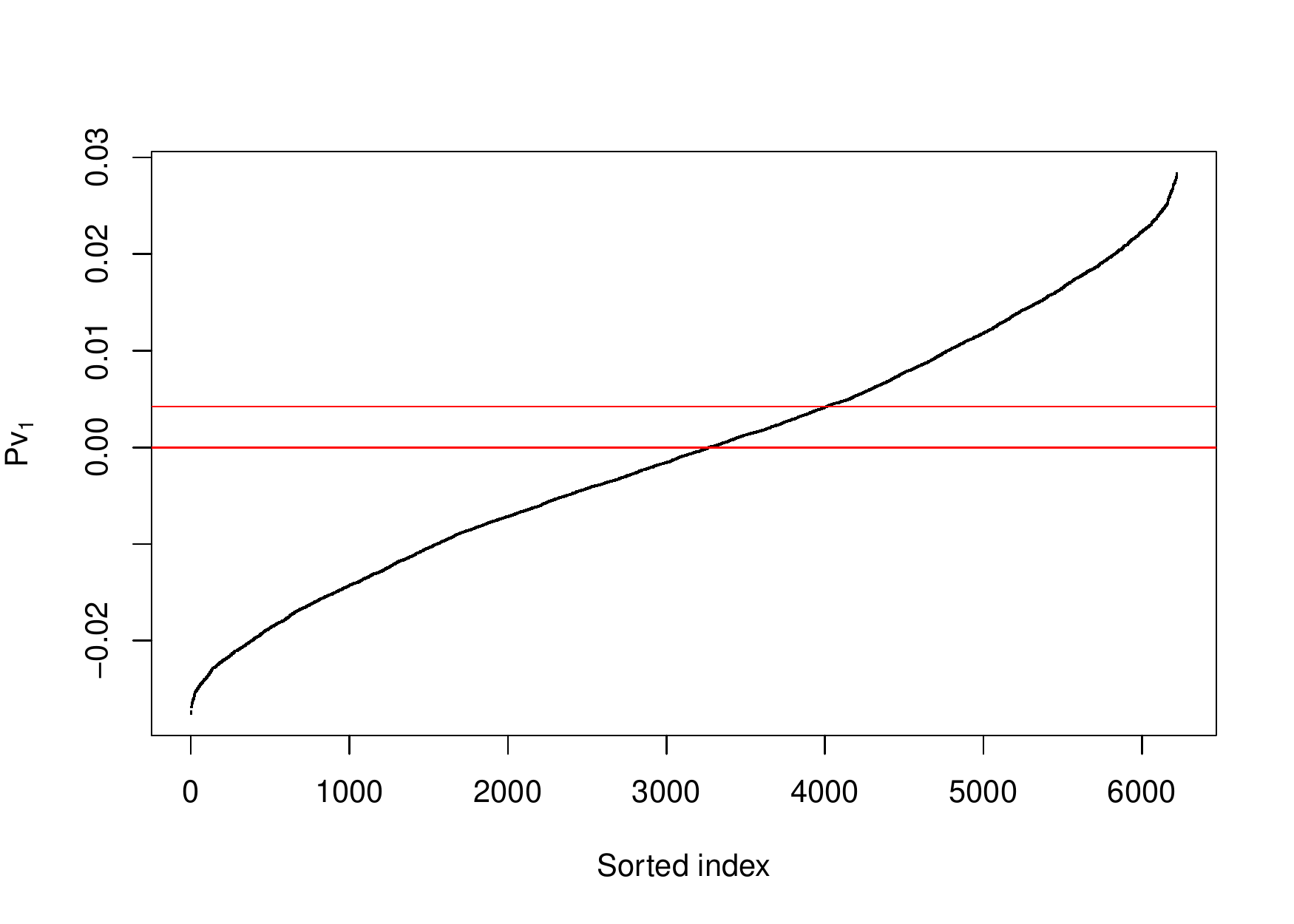}
\caption{
Example plot of the ordered entries of $v_1$ for the EisenYeast example from
Section \ref{examples}. The red lines illustrate the interval corresponding
threshold $t=0.99$ placed at an arbitrary location on the vertical axis. Points
farther apart than this interval correspond to column vectors that do not meet
the correlation threshold.
\label{fig1}
}
\end{center}
\end{figure}

Let $t$ be a given correlation threshold, $0<t<1$, and let
$P\in\mathcal{R}^{n\times n}$ be a permutation matrix that orders the entries
of $v_1$ in increasing order.  For example, Figure \ref{fig1} displays the
ordered entries of $Pv_1$ for the small ``EisenYeast'' example (where
$A\in\mathcal{R}^{80 \times 6221}$) from Section \ref{examples}.  The lines in
the plot illustrate the interval associated with the correlation threshold
$t=0.99$ placed at an arbitrary vertical axis location.

Let $D^i\in\mathcal{R}^{n-i, n}$, $i=1,2,\ldots,n-1$, be the $i$th order finite
difference matrix with $-1$ along the diagonal, $1$ along the $i$th
super-diagonal, and zeros elsewhere.  Then $D^1P v_1$ consists of the
differences of adjacent entries in $P v_1$.  Entries such that $\sigma_1^2 (D^1
P v_1)^2 > 2(1-t)$ correspond to pairs of vectors that do not meet the
correlation threshold, where here and below an exponent applied to a vector
denotes element-wise exponentiation.

Even a 1-d projected interval distance threshold may prune many possible
non-adjacent vector pairs with respect to the ordering defined by $P$ as
illustrated by the example in Figure \ref{fig1}.  However, this observation is
unlikely to rule out many adjacent pairs of vectors in most problems.  For
instance, the maximum distance between adjacent points shown in the example in
Figure \ref{fig1} is less than $\sqrt{2(1-t)}/\sigma_1$, and no pruning of
adjacent pairs of vectors with respect to the ordering $P$ occurs.

Including more terms from Equation \ref{vdist} increases our ability to prune
adjacent pairs of vectors below the correlation threshold with respect to the
ordering defined by $P$.  Including only one more term finds that 96\% of
adjacent vectors fall below the correlation threshold of $0.99$ in the
example shown in Figure \ref{fig1} and are pruned by identifying indices such
that
\[
\sigma_1^2 (D^1 P v_1)^2 +
\sigma_2^2 (D^1 P v_2)^2 > 2(1-t).
\]
Note that we use the permutation $P$ defined by the order of vector $v_1$
throughout.  In general, using $p\le k$ terms to prune pairs of vectors
below the correlation threshold boils down to evaluating the expression
\begin{equation}\label{proj}
(D^1 P V_{1:p} )^2 s_{1:p}^2 > 2(1-t),
\end{equation}
where 
$V_{1:p}$ denotes the first $p$ columns of the
matrix $V$, $s_{1:p}$ the vector of the first $p$ singular values.

Let $\ell$ be the longest run of successive points in $P v_1$ within the
interval $\sqrt{2(1-t)}/\sigma_1$.  The quantity $\ell$ may, for example, be
obtained by rolling the interval over all the points in Figure \ref{fig1} and
counting the maximum number of points that lie within the interval.  With
respect to the ordering $P$, $\ell$ represents the biggest index difference
that pairs of indices corresponding to correlated vectors above the threshold
can exhibit.

Pairs of vectors below the correlation threshold that lie more than one index
difference apart relative to the permutation $P$ can be similarly pruned by
replacing $D^1$ with $D^j$, $j=2,3,\ldots,\ell$ in the above expressions.
Following this process, we can produce a well-pruned candidate set of vector
pairs that contains all pairs that meet a given correlation threshold using
$\ell$ matrix vector products with matrices of order $n \times p$, where $p$ is
chosen in practice such that $p<<k$. Setting $p=10$ for the example shown in
Figure \ref{fig1} with $t=0.99$ prunes all but $149$ possible vector pairs out of a total
$6221(6221 - 1)/2 = 19,347,310$ with $\ell=787$.

These steps are formalized in Algorithm \ref{prune} below. The algorithm
proceeds in two main parts: steps 1--6 prune pairs of vectors down to a small
candidate set that may meet the correlation threshold; step 7
computes the correlation values and applies the threshold across the reduced
set of vector pairs.

\begin{algorithm}\label{prune}
Input: data matrix $A\in \mathcal{R}^{m\times n}$ with columns scaled to have
zero mean and unit norm, correlation threshold $t$,
truncated SVD rank $p<<k$.
\begin{enumerate}
\item Compute the truncated SVD
$AV_{1:p} = U_{1:p}\mathrm{diag}(s_{1:p})$,
where\\
$V_{1:p}=[v_1, v_2, \ldots, v_p]$
and $\mathrm{diag}(s_{1:p})$ denotes the diagonal matrix of the first $p$ singular values.
\item Compute permutation $P$ that orders points in $v_1$.
\item Compute $\ell$, the longest run of successive points in $P v_1$ within the interval $\sqrt{2(1-t)}/\sigma_1$.
\item Compute a set of candidate index pairs $(\tilde{j},\tilde{j}+\tilde{i})$ with respect to the index order defined by the permutation $P$ that possibly meet the correlation threshold
\[
\bigcup_{\tilde{i}=1}^\ell
\left\{
(\tilde{j},\tilde{j}+\tilde{i}) : 
e_{\tilde{j}}^T(D^{\tilde{i}} P V_{1:p} )^2 s_{1:p}^2 \le 2(1-t)
\right\},
\]
where $\tilde{j}=1,2,\ldots,n-\tilde{i}$.
\item If there are judged to be  ``too many'' candidate pairs, increase $p$,
compute $AV_{1:p} = U_{1:p}\mathrm{diag}(s_{1:p})$ and go to step 4.
Continue to increase $p$ in this way until the number of candidate
pairs is sufficiently small or stops decreasing (the best this algorithm can do).
\item Recover the non-permuted candidate column vector
indices $(j,i)$ from $j=e_{\tilde{j}}^T Pq$ and $i=e_{\tilde{j}+\tilde{i}}^TPq$ for every
pair $(\tilde{j},\tilde{j} + \tilde{i})$ in step 4, where
$q\in\mathcal{R}^n = [1,2,\ldots,n]^T$.
\item Compute the full correlation coefficient for each column vector pair identified in
the previous step, applying the correlation threshold to this reduced set of values.
\end{enumerate}
\end{algorithm}
One can cheaply estimate whether there are ``too many'' candidate pairs for a
given dimension $p$ in step 5 by evaluating step 4 for only adjacent pairs of
vectors, $\tilde{i}=1$.

Algorithm \ref{prune} guarantees that no pruned vector pair exceeds the given
correlation threshold--the pruning does not produce false negatives.  The
algorithm typically prunes the vast majority of pairs of vectors below the
correlation threshold with truncated singular value decompositions of only
relatively low rank $p$.  The choice of $p$ represents a balance between work
in computing a truncated SVD and evaluating $\ell$ rank $p$ matrix products in
Step 4 (each increasing in work as $p$ increases), and the work required to
filter the thresholded values in step 7 (decreasing work as $p$ increases).

Most of the floating point operations occur in step 4 of Algorithm \ref{prune}.
Fortunately the $\ell$ evaluations in step 4 are independent of each other and
can easily be computed in parallel. See the reference \proglang{R}
implementation \cite{sup} for an example.

\section{General matrices and fast truncated SVD}\label{irlba}

We have so far assumed that the mean of each column vector of the $m\times n$
matrix $A$ is zero and the columns are scaled to have unit norm, but we really
want a method that works with general matrices $A$.  We also want a way to
efficiently compute a relatively low-rank truncated SVD of potentially large
matrices required by step 1 of Algorithm \ref{prune}.  And the truncated SVD
method should be able to cheaply restart to compute additional singular vectors
as required in step 5 of Algorithm \ref{prune}. Fortunately all desires are met
by one algorithm.

The augmented implicitly restarted Lanczos bidiagonalization algorithm\break(IRLBA)
of Baglama and Reichel \cite{irlba} efficiently computes truncated singular value
decompositions of large dense or sparse matrices. Reference implementations are
available for \proglang{R} \cite{irlbar}, \proglang{Matlab} \cite{irlbam},
and \proglang{Python} \cite{irlbap}.

We can relax the column mean and scale assumptions without introducing much
additional computational or storage overhead as follows. Assume $A$ is an
$m\times n$ real-valued matrix without any constant-valued columns. Let
$z=[z_1,z_2,\ldots,z_n]\in\mathcal{R}^n$ represent the vector of column means
of $A$, $w=[w_1^2,w_2^2,\ldots,w_n^2]\in\mathcal{R}^n$
be the vector of squared column norms of $A$, and $W\in\mathcal{R}^{n\times n}$
a diagonal matrix with diagonal entries $1/\sqrt{w_i^2 - m z_i^2}$.  Let
$e\in\mathcal{R}^m$ be a vector of all ones.  Then $(A - ez^T)W$ is a centered
matrix with zero column means and scaled to have unit column norms, and
\[
\mathrm{cor}(A) = W^T (A-ez^T)^T (A-ez^T) W.
\]
The IRLBA, based on the Lanczos process, is a method of iterated matrix-vector
products. The main idea behind efficient application of IRLB to correlation
problems replaces matrix-vector products of the form $Ax$ with $AWx - ez^TWx$ in
the iterations, implicitly working with a scaled and centered matrix without
forming it.  This comes at the cost of storing two additional length $n$ vectors and
at most computing two additional vector inner products per matrix-vector product.
The \proglang{R} implementation \cite{irlbar} of IRLBA includes arguments for centering and
scaling the input matrix.

A na\"ive brute force thresholded correlation computation requires $O(n^2 m)$ flops.
Let $p$ be the selected IRLBA dimension. Then Algorithm \ref{prune} requires
$O(n p \ell)$ flops, where $\ell$ is the longest run of ordered entries in
$v_1$ that meet the 1-d projected distance threshold $\sqrt{2(1-t)}/\sigma_1$,
including the flops required by IRLBA and final thresholding in step 7.  The
main computational savings of $\frac{nm}{lp}$ depends on the rate of decay in
the singular values of $A$, the subspace dimension $p$, and the desired
threshold value. We remark that this estimate of computational savings is
available already in step 3 of Algorithm \ref{prune} after only modest
computational effort. Users might use this cheap estimate to, for example,
alter their threshold in some cases.

\section{Numerical experiments}\label{examples}

We evaluated the algorithm using public gene expression and DNA methylation
datasets from the \proglang{R} {\tt biclust} \cite{biclust} package and from
the Cancer Genome Atlas \cite{gdac}, referred to as TCGA.  Most tests were
performed on a desktop PC with a single AMD A10-7850K 3.7$\,$GHz Athlon
quad-core CPU and 16$\,$GB 1333$\,$MHz unbuffered RAM running the Ubuntu 15.10
GNU/Linux OS. Tests were performed with 64-bit \proglang{R} version 3.2.3 using
double-precision arithmetic and the OpenBLAS library version 0.2.14-1ubuntu1
(based on Goto's BLAS version 1.13) with {\tt OMP\_NUM\_THREADS=1}. The last
experiment includes a test run on a Linux cluster described in that section.

All examples use the reference \proglang{R} implementation, {\tt tcor()}, which can be
installed from the development GitHub repository using the
{\tt devtools} package with:

\begin{Schunk}
\begin{Sinput}
> devtools::install_github("bwlewis/tcor")
\end{Sinput}
\end{Schunk}

The native \proglang{R} {\tt cor()} function is a general function that can
compute several different types of correlation matrices including Pearson,
Spearman, and Kendall and includes a number of options for dealing with missing
values.  Because of its flexibility, the native {\tt cor()} function does not
always compute Pearson correlation matrices in the most
computationally-efficient way.

Correlation matrix computation can benefit substantially from use of optimized
BLAS Level 3 operations (matrix multiplication), achieving high utilization of
available CPU floating point capacity. The {\tt tcor()} function also benefits
from optimized BLAS routines, but not as much, as it mostly makes use of Level
2 operations and Level 3 operations on sequences of smaller problems.

We compare the {\tt tcor()} function with brute-force Pearson correlation
matrix computation computed in the fastest way we can, directly using matrix
multiplication instead of using the more general {\tt cor()} function.

\subsection{Small gene expression example}

The first example is a small proof of concept that provides the data for Figure
\ref{fig1} and establishes our timing and memory use measurement methodology.
Wall-clock times are reported in seconds and peak memory use in excess of
memory required to store the input data matrix is reported in megabytes.

Anyone can quickly run the example to experiment with the algorithm and
compare its output with brute force results. The example uses the {\tt
EisenYeast} data from the {\tt biclust} \proglang{R} package \cite{biclust}.
The data consist of 80 experiments involving the
Saccharomyces Cerevisiae yeast measuring gene expression data across 6,221
Affymetrix probes, represented as an $80\times 6221$ real-valued matrix.  The
example finds all pairs of columns (gene expression measurements) that meet or exceed a
Pearson correlation value of 0.95. The choice of projection dimension $p$ affects
the efficiency of the algorithm. For $p=2$ the number of candidate vector pairs
generated in step 4 of Algorithm \ref{prune} is over 3 million. When $p$ is increased
to $10$, the number of candidate vector pairs drops to less than 13,000.
The reference \proglang{R} implementation includes options for
tuning $p$ and relies on the restart ability of the IRLBA to pick up where it left off and
inexpensively extend an existing truncated SVD. The following example uses $p=10$.
\begin{small}
\begin{Schunk}
\begin{Sinput}
> threshold = 0.95
> # Test system values:
> t1=3.13
> t2=1.75
> m1=59.60
> m2=1075.10
> if(Sys.getenv("RUN_EXAMPLES")=="TRUE")
+ {
+ library(tcor)
+ library(doMC)     # local parallel computation
+ library(biclust)  # for the BicatYeast data
+ data(EisenYeast)  # in gene by sample orientation
+ A = t(EisenYeast) # to correlate genes to genes
+ # Register a four-core multicore parallel backend
+ registerDoMC(4)
+ 
+ # Compute using tcor():
+ m1  = sum(gc()[,2])  # Memory in megabytes currently in use
+ t1  = proc.time()
+ tx  = tcor(A, t=threshold, p=10)
+ t1  = (proc.time() - t1)[3] # Time in seconds
+ m1  = sum(gc()[,6]) - m1    # Peak excess memory use during the test
+ 
+ # Fast brute force full correlation:
+ m2  = sum(gc()[,2])
+ t2  = proc.time()
+ mu = colMeans(A)
+ s  = sqrt(apply(A, 2, crossprod) - nrow(A) * mu ^ 2)
+ cs = scale(A, center=mu, scale=s)  # scale and center the data
+ cx = crossprod(cs)                 # full correlation matrix
+ cx  = cx * upper.tri(cx)    # ignore symmetry and diagonal
+ pairs = which(cx >= threshold, arr.ind=TRUE)
+ t2  = (proc.time() - t2)[3]
+ m2  = sum(gc()[,6]) - m2
+ }
\end{Sinput}
\end{Schunk}
\end{small}
Brute-force computation outperforms {\tt tcor()} in this small first example in
speed but consumes substantially more memory.  Each algorithm identifies the
same set of 125 gene expression pairs that exceed the specified correlation
threshold of 0.95, with comparative timing and memory use shown in the
following table.

\begin{table}[ht]
\centering
\begingroup\small
\begin{tabular}{lrr}
  \hline
Method & Wall clock time (s) & Peak excess memory use (MB) \\ 
  \hline
tcor & 3.13 & 59.60 \\ 
  brute force & 1.75 & 1075.10 \\ 
   \hline
\end{tabular}
\endgroup
\caption{Gene expression results, threshold=0.95} 
\label{EisenYeast}
\end{table}

\subsection{TCGA DNA methylation example}

Performance benefits from the {\tt tcor()} function become apparent with more
data.  The following example uses calculated beta values from the Illumina
Human Methylation450 BeadChip of 80 Adenoid Cystic Carcinoma (ACC) tumor
samples obtained from the Broad GDAC Cancer Genome Atlas dashboard \cite{gdac}.
See the \proglang{R} package implementation \cite{sup} for examples of
efficiently converting the downloaded sample methylation data into an R matrix.
We chose these data for their public availability as a general example of
computing thresholded correlation among large numbers of vectors,
rather than for any specific biological meaning.

We removed 91,563 columns with all constant or missing values from the data,
leaving an $80\times 394014$ matrix. The example computes pairs of methylation
beta value column vectors with correlation values of 0.99 or more. We used a
{\tt tcor} projection dimension of 10 and the same performance measurement
methodology used in the first example. Despite its low rank and modest data
size, the problem is too large to compute using a na\"ive brute force
correlation method.  (We remark that the matrix multiplication central to the
brute force implementation can be decomposed into a sequence of smaller steps
to fit within given memory constraints. However, the brute force algorithm
would still need to evaluate all possible vector pair
correlation values.)

Algorithm \ref{prune} identified 913,601 pairs of vectors with correlation
values of 0.99 or more in about three hours on our test PC. A brute force
algorithm would need to evaluate the threshold condition for more than 77
billion vector pairs; the projection of Algorithm \ref{prune} limited the
search space to only about 4 million candidate pairs, from which the 913,601
correlated pairs were found.

We illustrate that steps 4--7 of Algorithm \ref{prune} can be computed in
parallel by comparing the above results run on our reference test PC with
results run on a modest GNU/Linux cluster at Paradigm4, Inc. The cluster
consisted of four computers connected by 10 gigabit ethernet. Each computer
contained two Intel Xeon E5-2650 2$\,$GHz CPUs (16 physical CPU cores per
computer) and 64$\,$GB/1,600$\,$MHz ECC registered RAM running CentOS 6.6 and
64-bit \proglang{R} version 3.2.2 with the reference \proglang{R} BLAS library.

No code change was required to run in parallel on the cluster. The reference \proglang{R}
{\tt tcor()} implementation uses \proglang{R}'s {\tt foreach} \cite{foreach} framework to
run on arbitrary parallel configurations (``back ends''). We simply registered
a {\tt doRedis} \cite{doredis} parallel back end with 16 \proglang{R} workers per computer
(64 total \proglang{R} worker processes) prior to invoking {\tt tcor()}.

The results computed on the Linux cluster are marked ``tcor (64 core cluster)''
below. The cluster example computed the filtering step 7 of Algorithm
\ref{prune} in parallel across the 64 remote \proglang{R} processes. That approach incurs,
in this example, an additional memory overhead of about 16$\,$GB because each
remote \proglang{R} process requires access to a copy of the data matrix $A$. More thrifty
adaptations of the algorithm are possible and could, for example, share the
memory required by $A$ between worker \proglang{R} processes on each computer.

\begin{table}[ht]
\centering
\begingroup\small
\begin{tabular}{lrrl}
  \hline
Method & Wall clock time (s) & Memory use (MB) &   \\ 
  \hline
tcor (4 core PC) & 10309.00 & 389.10 &  \\ 
  tcor (64 core cluster) & 482.00 & 994.30 & (local) \\ 
   &  & 42282.40 & (remote) \\ 
   \hline
\end{tabular}
\endgroup
\caption{TCGA ACC Methylation450 (80 rows x 394,014 columns, threshold=0.99). Linux cluster results include master R process memory use (local) plus total peak memory used by all 64 worker R processes (remote).} 
\label{TCGA3}
\end{table}From a set of almost 80 billion pairs of vectors, Algorithm \ref{prune} finds 913,601
pairs with correlation values of 0.99 or more on a modest Linux
cluster in about 8 minutes. The same algorithm works, more slowly, in
the constrained memory setting of a desktop PC without modification.

\section{Extensions}\label{extensions}

The idea of evaluating threshold conditions by projection into low dimensional
subspaces applies to distance metrics similarly to correlation, see
\cite{svd-similarity} for a related approach. Lemma \ref{sumlemma} decomposes
Euclidean distance between two vectors in terms of singular basis vectors and
Algorithm \ref{prune} can be applied directly to the problem of finding all
pairs of vectors that meet a thresholded Euclidean distance with small changes
in steps 3, 4, and 7.  By equivalence of norms, thresholded Euclidean distance
can be extended to other distance metrics. The reference \proglang{R}
implementation \cite{sup} includes functions for thresholded distance and
correlation.

The ideas presented here can be adapted to efficiently find the top N most
correlated vector pairs for a given constant N similarly to the work of
Xiong et al. \cite{prune2}.

\section{Conclusion}\label{conclusion}

We present a simple algorithm and reference \proglang{R} implementation based
on a truncated singular value decomposition (SVD) that efficiently computes
thresholded subsets of large correlation  matrices.  We show that the natural
connection between data covariance and the SVD can result in substantial
computational and memory savings, and illustrate these savings with a few
genomics data experiments.  The main computational work in the algorithm is
easily parallelized and deployed across computational clusters.

\bibliography{bibliography}{}
\bibliographystyle{plain}

\end{document}